\documentclass[a4paper, amsfonts, amssymb, amsmath, reprint, showkeys, nofootinbib, twoside,superscriptaddress, prb, pdflatex]{revtex4-1}
\usepackage{graphicx}
\begin{document}
\title{Formation of graphene nanoribbons on the macrofacets of vicinal $6H$-SiC(0001) surfaces}

\author{Kohei Fukuma}
\author{Anton Visikovskiy}
    \affiliation{Department of Applied Quantum Physics and Nuclear Engineering, Kyushu University, Fukuoka 819-0395, Japan}
\author{Takushi Iimori}
\author{Fumio Komori}
    \affiliation{Institute for Solid State Physics, The University of Tokyo, Kashiwa, Chiba 277-8581, Japan}
\author{Satoru Tanaka}
    \affiliation{Department of Applied Quantum Physics and Nuclear Engineering, Kyushu University, Fukuoka 819-0395, Japan}
    \email[Correspondence email address: ]{stanaka@nucl.kyushu-u.ac.jp}

\date{\today} 

\begin{abstract}
Thermal decomposition of vicinal $6H$-SiC(0001) surfaces with off-angles toward the $[1\bar{1}00]$ direction results in the appearance of pairs of $(0001)$ macroterraces and $(1\bar{1}0n)$ macrofacets covered with graphene, as follows. A carpet-like carbon layer grows on the surface, covering both the macroterraces and macrofacets; it forms $(6\sqrt{3} \times 6\sqrt{3})$ buffer layer on the former ones, whereas its partial periodic bonding with the SiC steps on the latter ones generates  a pseudo-graphene nanoribbon (pseudo-GNR) array. The nanoribbons have a width of 2~nm and are aligned in the $[1\bar{2}10]$ direction with a spatial periodicity of 3.3~nm. Here, the Raman spectroscopy analysis of the pseudo-GNR array showed the absence of the 2D peak and the polarization dependence of the $G$ and $D$ peaks, which is typical of the armchair edge nanoribbon.
\end{abstract}


\maketitle

\section{Introduction}\label{sec:intro}
Graphene is a two-dimensional material with exotic electronic properties and very high carrier mobility, which makes it an interesting candidate for next-generation electronics \cite{a1,a2}. Application in traditional logic switching devices, however, requires the existence of a band gap that normal graphene lacks. Therefore, methods for opening a gap in the graphene band structure have been investigated. For example, cutting graphene into graphene nanoribbons (GNRs) with armchair edges can generate a substantial energy gap, but the ribbon width must be accurately controlled \cite{a3,a4}. Lithography\cite{b1} and unzipping carbon nanotubes\cite{b2} are traditional top-down approaches to fabricate GNRs, while bottom-up approaches include GNR synthesis from molecular precursors\cite{b3,b4}. Both strategies present some drawbacks, though; the top-down approaches struggle with accurate control of width and edge quality, while the bottom-up ones require specific metal substrates (i.e., the requirement of post-growth transfer procedure) and suffer in the ribbon alignment.

Today, GNR growth on semiconductor SiC substrates is attracting attention as a new bottom-up method. Through molecular beam epitaxy, an array of GNRs having ${\sim}5$~nm width can be grown on the nanoperiodic structure \cite{c1,c2} formed on a vicinal SiC substrate \cite{c3}; the resulting GNRs are periodic with ${\sim}10$~nm spatial interval and exhibit a band gap of 0.14~eV, as observed via angle-resolved photoemission spectroscopy (ARPES). Graphene grown on a trench or mesa structures premade on SiC(0001) surfaces has also been studied\cite{d1,d2,d3,d4,d6,d5}. The edge structure of GNRs grown on trench sidewall is controlled by trench orientation; specifically, the GNRs having zigzag edges grown on the sidewall along the $[11\bar{2}0]$ direction of $6H$-SiC(0001) surfaces show the characteristics of ballistic conduction \cite{d1,d2}. Furthermore, both the zigzag-edge GNRs grown on the sidewall of $4H$-SiC(0001) surfaces\cite{d3} and the armchair-edge ones grown on $6H$-SiC(0001) surfaces\cite{d4} exhibit band gaps.

We previously have studied the formation and growth mechanism of graphene on terrace-facet periodic structures (reported as macrostep formation on vicinal SiC(0001) surfaces\cite{e1,e2}). Step bunching during thermal decomposition results in the formation of macroterraces and macrofacets with a period of several hundred nanomaters. Ienaga \textit{et al.} \cite{f1} investigated bilayer graphene formed on macrofacets and observed the periodic modulation of electron-phonon coupling in the top graphene layer via scanning tunneling spectroscopy. This result suggests that the bottom carbon layer is not continuous graphene but an array of GNRs.

In the present work, we grew uniform single graphitic layer on macrofacets (hereafter defined as facet-graphene) and studied its structure and physical properties. This facet-graphene was attached periodically to the underlying substrate structures, and this repetition of freestanding portions of the carbon layer and those bonded to the substrate formed a pseudo-GNR array. Its Raman spectrum revealed features that are characteristic of armchair-edge GNRs and a width of several nanomaters.

\section{Experimental}\label{sec:methods}
A $6H$-SiC(0001) substrate with a $15^{\circ}$ miscut angle toward the $[1\bar{1}00]$ direction was used to grow the GNR array. Such a large miscut angle was used to ensure a large facet/terrace area ratio, which is important for surace characterization via  macroprobe methods such as low-energy electron diffraction (LEED) and ARPES; besides, the resulting low facet angle with respect to the surface is more accessible for scanning probe methods. First, the samples were cleaned and etched via high temperature annealing ($1360^{\circ}$C) in a hydrogen flow under atmospheric pressure. After confirming through atomic force microscopy (AFM) the surface morphology and the absence of polishing damage, we proceeded with surface thermal decomposition in an Ar atmosphere at ambient pressure and $1500^{\circ}$C; this treatment led to step bunching, resulting in a periodic surface morphology consisting of (0001) macroterraces and macrofacets (typically $27^{\circ}$-inclined with respect to (0001)). The samples were then analyzed by AFM, LEED, scanning tunneling microscopy (STM), and microRaman spectroscopy.

\section{Results and Discussion}\label{sec:results}
\begin{figure}
    \centering
    \includegraphics[width=\columnwidth]{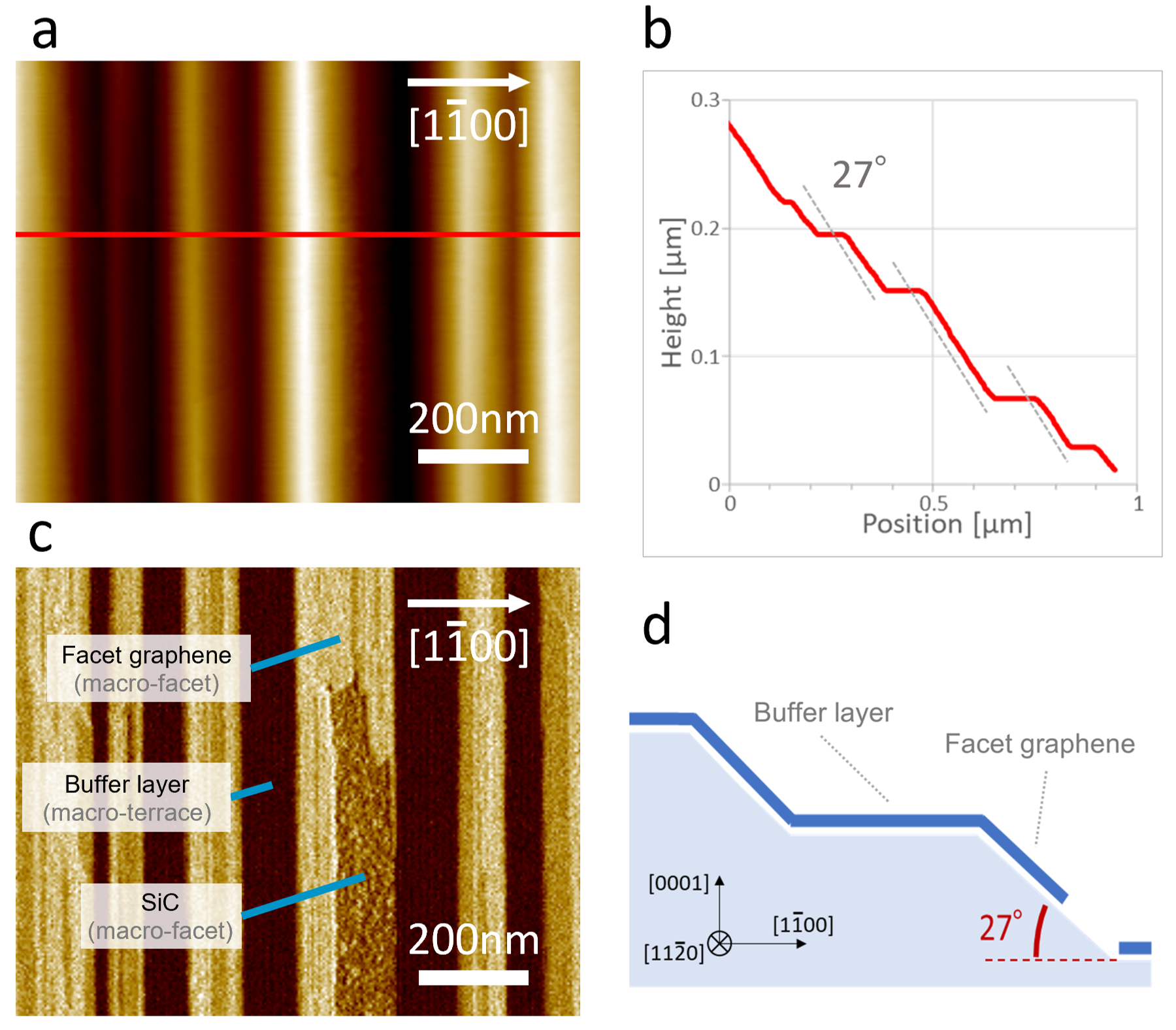}
    \caption{
    Atomic force microscopy results.
    (a) Height and (c) phase images of the sample after surface decomposition.
    (b) Cross-sectional height profile along the red line in (a); the gray dotted line represents the $27^{\circ}$ angle from the $(0001)$ plane.
    (d) a schematic diagram of the macroterrace-macrofacet structure after facet-graphene growth.}
    \label{fig:fig1}
\end{figure}

The AFM observation of the sample surace after thermal decomposition (Fig~\ref{fig:fig1}(a)) showed pairs of macroterraces and macrofacets, which had individual widths ranging from tens to ${\sim}200$~nm, along the $[1\bar{1}00]$ direction. The corresponding cross-sectional profile (Fig.~\ref{fig:fig1}(b)) revealed that the macrofacets formed an angle ${\sim}27^{\circ}$ with respect to the (0001) macroterraces.

In the AFM phase image (Fig.~\ref{fig:fig1}(c)) of the same area, three types of regions of different contract could be clearly distinguished: dark areas on the macroterraces represent insulating $(6\sqrt{3} \times 6\sqrt{3})R30^{\circ}$ buffer layer (hereafter denoted as 6R3)\cite{g1,g2}, and bright and grainy mid-tone areas, respectively paved and unpaved by graphene, on the macrofacets. The facet $27^{\circ}$ angle, as shown in cross-section (Fig.~\ref{fig:fig1}(d)) is related to the most energetically stable configuration, as reported in previous studies\cite{d6,f1}. We will focus on the facet structures in the following paragraphs.

\begin{figure}
    \centering
    \includegraphics[width=\columnwidth]{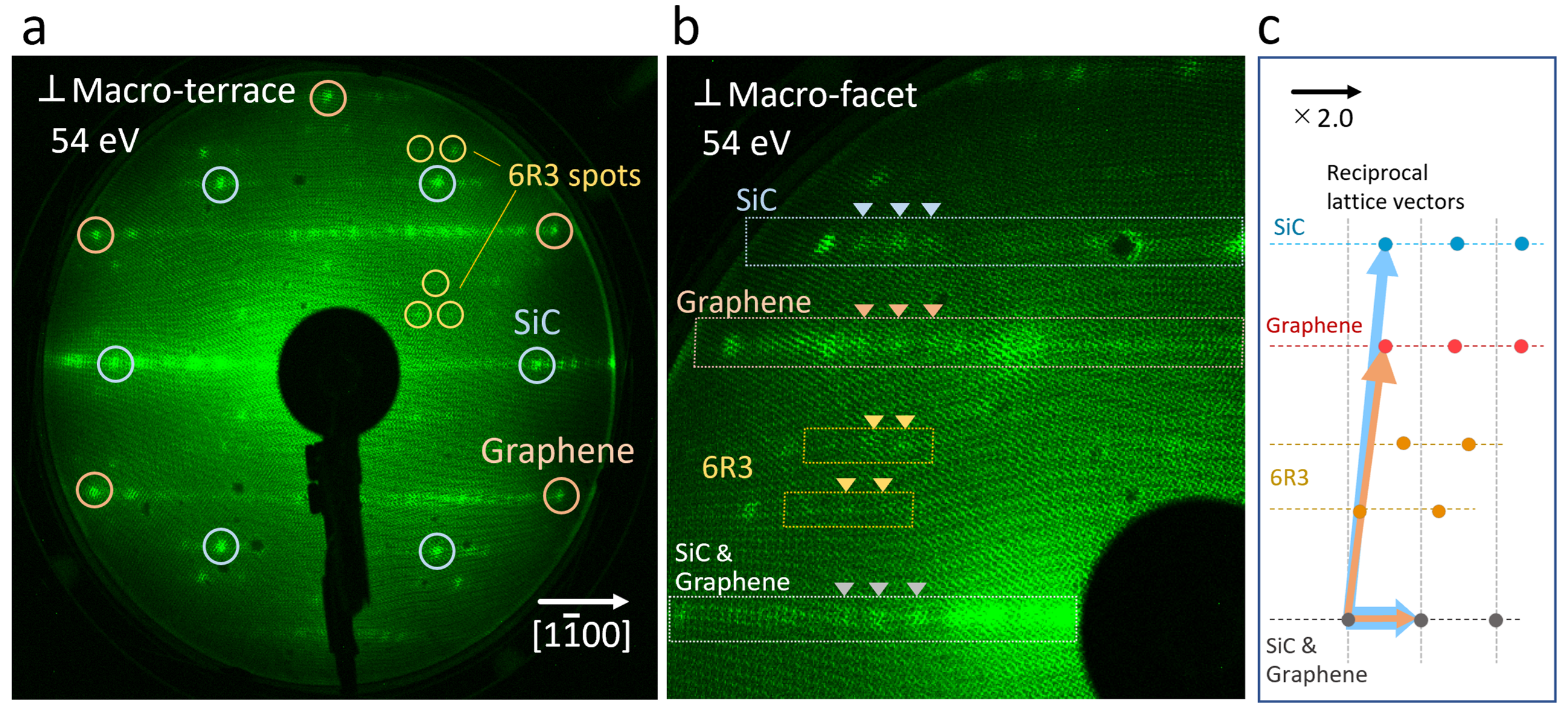}
    \caption{Low-energy electron diffraction patterns, obtained with a beam incidence normal to (a) the macroterraces and (b) the macrofacets, with an electron energy of $54$~eV. (c) Correspondence between some of the super spots shown in (b), with triangular marks, and SiC, graphene, and the 6R3 structure superlattice; the horizontal scale is doubled.}
    \label{fig:fig2}
\end{figure}

The LEED measurements were performed at two different incident beam directions, that is, normal to the macroterraces and macrofacets. In the first case (Fig.~\ref{fig:fig2}(a)), besides the typical $(1,0)$ spots of epitaxial graphene and SiC(0001) substrate and the satellite spots of the 6R3 buffer layer, chain like super-spots aligned along the $[1\bar{1}00]$ direction were observed. When the electron energy was changed, these super-spots moved not only toward the center of the LEED pattern, as other major spots did, but also slid along the $[1\bar{1}00]$ direction; this indicates that they came from the macrofacets and not from the macroterrace area. Indeed, when the sample was rotated $27^{\circ}$, that is, when the beam was normal to the macrofacets, the movement of these super-spots with varying of the electron energy became focused toward the pattern center (Fig.~\ref{fig:fig2}(b)). In that case, the chain-like super-spots were observed around not only the main graphene spots but also the SiC and 6R3 reflexes. As will be discussed below, this was due to the moir\'{e} pattern between facet-graphene and SiC lattice along the  $[11\bar{2}0]$ direction. All the super-spots were equidistant and indicated a real-space periodicity of ${\sim}3.3$~nm. Fig.~\ref{fig:fig2}(c) shows a diagram of the reciprocal cells corresponding to the observed periodic structures on the macrofacets.

\begin{figure*}
    \centering
    \includegraphics[width=0.8\textwidth]{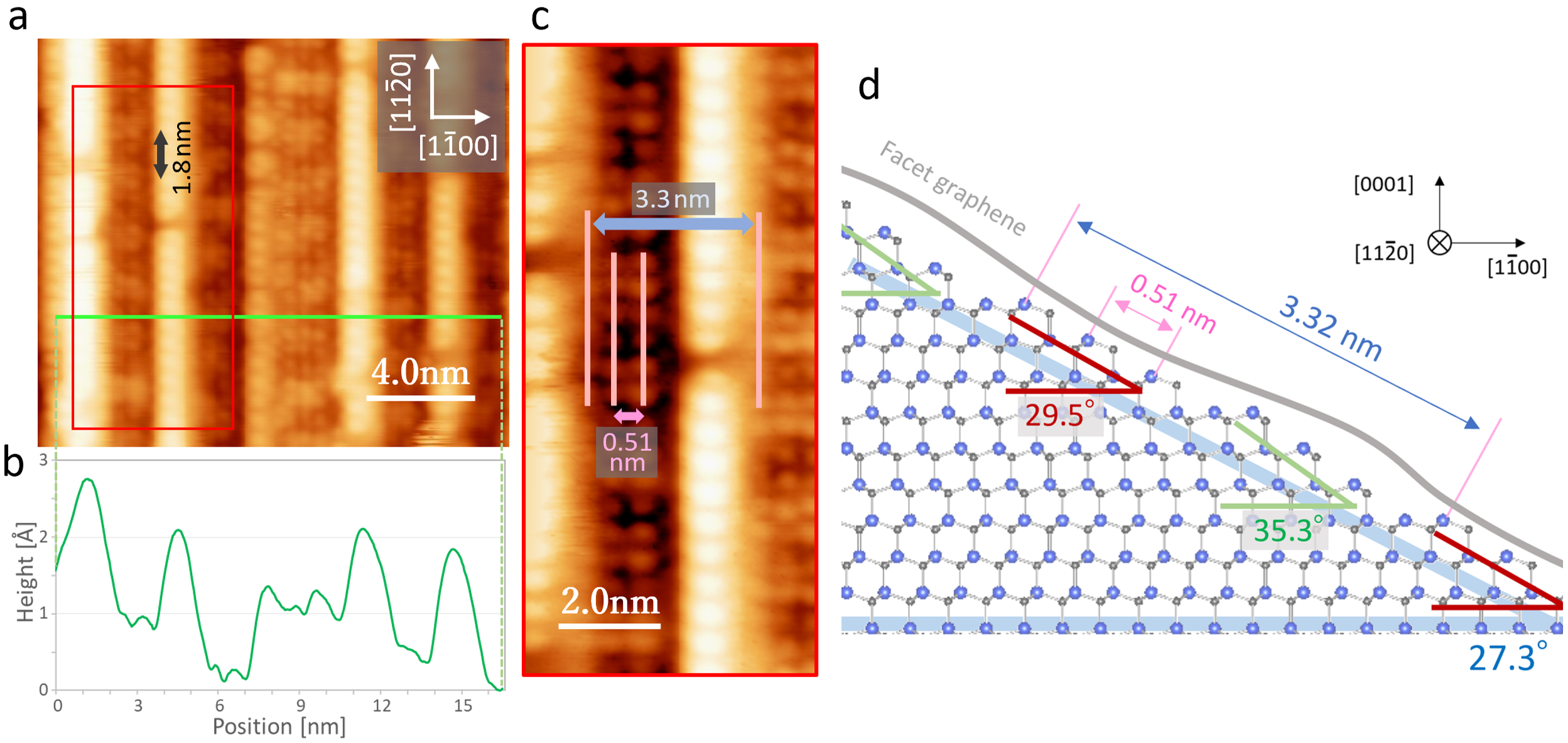}
    \caption{(a) Scanning tunneling microscopy (STM) image of facet-graphene on the macrofacets. (b) Cross-sectional height profile along the green line in (a). (c) STM image taken in the red frame in (a). (d) Cross-sectional model of the macrofacets.}
    \label{fig:fig3}
\end{figure*}

Then, the macrofacets were studied by STM. An STM topographic image measured at a bias voltage of  $-1.5$~V (filled states) showed a bright and dark striped pattern along the  $[1\bar{1}00]$ direction (Fig.~\ref{fig:fig3}(a)). As shown by its cross-sectional height profile (Fig.~\ref{fig:fig3}(b)), the overall periodicity was $3.3$~nm, which coincides with the calculated spacing of the SiC LEED super-spots shown in (Fig.~\ref{fig:fig2}). The valley (darker) regions exhibited a fine structure with dot-like features lining up in the $[11\bar{2}0]$ direction at an interval of ${\sim}1.8$~nm, which is about six times the SiC lattice constant and coincides with the period of the 6R3 structure on SiC(0001) surfaces in the $[11\bar{2}0]$ direction \cite{LEED-1}. Therefore, we assume that this is a one-dimensional moir\'e pattern between graphene and SiC lattice on macrofacets, similar to the 6R3 buffer layer. The LEED super-spots being lined up along the 6R3 spots in Fig.~\ref{fig:fig2}(b) also support this feature. Fig.\ref{fig:fig3}(c) shows the STM topographic image of the area outlined by the red frame in (a), taken at a lower bias voltage ($-0.5$~V), where the dot-like features in a valley can be observed more clearly; they formed a triple raw with a line spacing of ${\sim}0.51$~nm, whose origin is discussed below.

Based on the AFM and STM results, we built a structure model of the macrofacet surface, having an angle of $27.3^{\circ}$ (Fig.~\ref{fig:fig3}(d)), which consists of miniterraces paired with $35.3^{\circ}$ and $29.5^{\circ}$ minifacets, corresponding to the ridges and valleys, respectively, in the STM image. The overall period is $3.32$~nm. These two characteristic minifacet angles arose because in  $6H$-SiC(0001) a stacking sequence of Si-C bilayers is switched every three of them, resulting in locally different facet angles relative to (0001). In the $29.5^{\circ}$ minifacets, each row of carbon and silicon atoms is separated by 0.51-nm intervals, which is consistent with the dot-raw spacing observed in the STM image in Fig.~\ref{fig:fig3}(c).

We further investigated the stability of each minifacet under monolayer graphene. The $29.5^{\circ}$ and $35.3^{\circ}$ minifacets contain two and one dangling bonds per carbon atom, respectively, and thus are correspondingly more and less reactive to graphene. As a result, the graphene on each minifacet can exhibit different characteristics in both bonding configurations and electronic structure; these differences were indeed highlighted by the STM image shown in Fig.~\ref{fig:fig3}(c). A buffer-like graphitic layer was formed on the $29.5^{\circ}$ minifacets while quasi-freestanding graphene, in the form of nanoribbons, grew on the $35.3^{\circ}$ ones. In the following discussion, we define the bright and dark regions in Fig.~\ref{fig:fig3}(a) as pseudo-GNRs and buffer layer ribbons (BLRs), respectively; the term pseudo-GNRs (hereafter abbreviated just as GNRs), which continuously connect to BLRs, is used in contrast to isolated GNRs.

\begin{figure}
    \centering
    \includegraphics[width=\columnwidth]{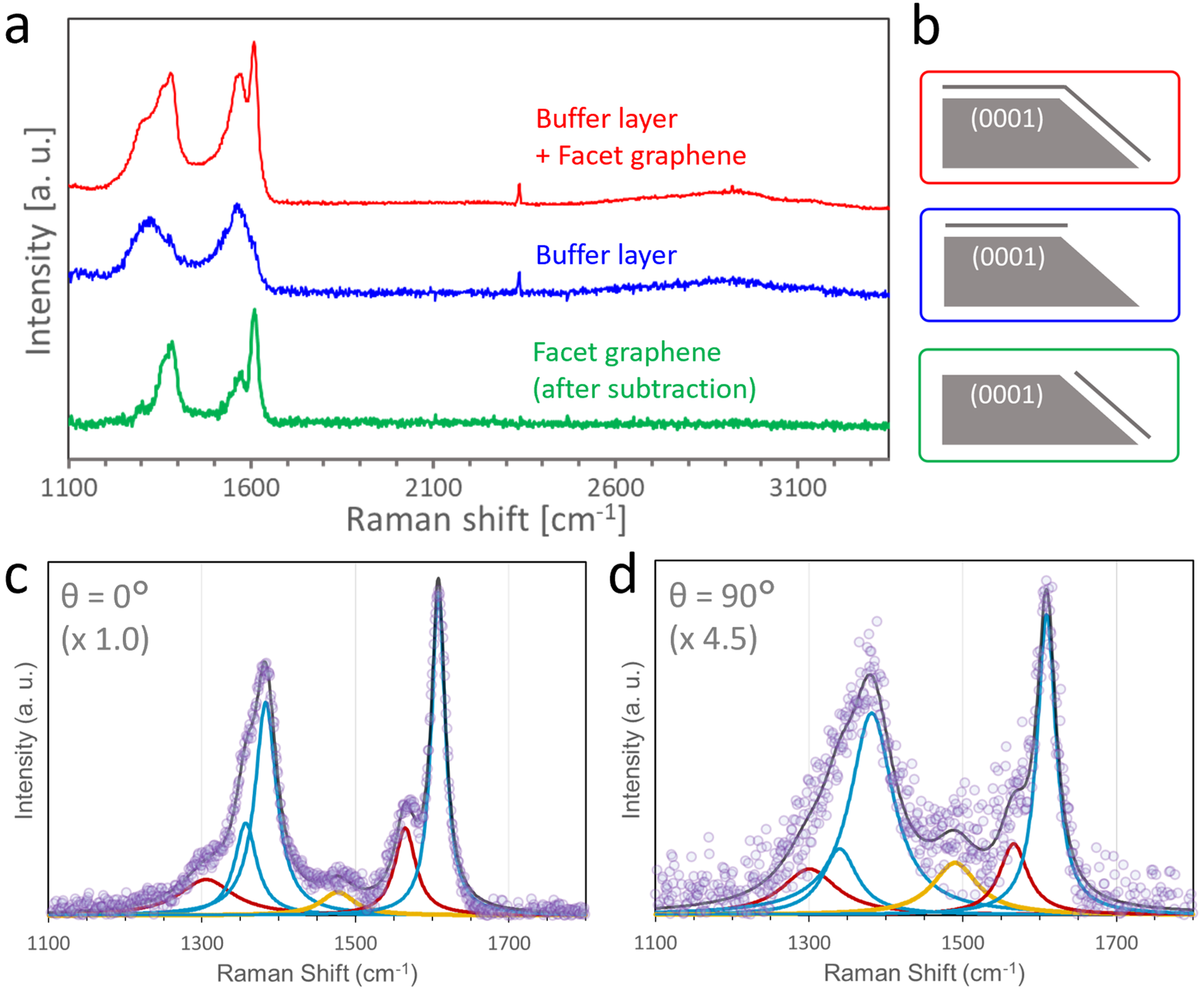}
  \caption{(a) Raman spectra of the samples covered entirely by a carbon layer and a buffer layer on the macroterraces only. (b) Cross-sectional views of the sample whose spectra are shown in (a). (c,d) Extracted Raman spectra of facet-graphene, measured with light polarization parallel (c) and perpendicular (d) to the graphene nanoribbon (GNR) edge; the lilac circles represent the measured values, the blue and red peaks correspond to the GNR and buffer layer ribbon regions in facet-graphene, respectively, and the solid black line is the total fit curve (the origin of the yellow peak is still unclear).}
    \label{fig:fig5}
\end{figure}

The samples were successively investigated via micro-Raman spectroscopy. Given the difficulty in extracting the Raman spectrum of the facet-graphene alone due to the limited spatial resolution (${\sim}1$ $\mu$m) of the equipment used, we prepared and analyzed two samples: one with a carbon layer covering the whole surface, i.e., both the macroterraces (in the form of the buffer layer\cite{Ra-a1,Ra-a2}) and macrofacets, and the other with a carbon layer only on the macroterraces (obtained at a reduced growth time) (Fig.~\ref{fig:fig5}(b)). Fig.~\ref{fig:fig5}(a) shows the measured spectra of both samples and the one of the facet-graphene alone, obtained by subtracting the blue spectrum from the red one. A Raman spectrum of graphene usually shows the $D$, $G$, and $2D$ peaks. In this study, the $2D$ peak, which is originated from the double resonance two-phonon inelastic scattering\cite{Ra-a5}, was absent. In the case of a narrow GNR, however, double inelastic scattering is much less probable than elastic scattering by the edges, rendering $2D$ peak almost negligible and $D$ peak relatively high. This phenomenon has also been observed in other GNRs \cite{Ra-a3,Ra-a4}. Thus, the absence of the $2D$ peak suggests the presence of narrow GNRs. In the $D$ and $G$ regions, three peaks components could be recognized (Fig.~\ref{fig:fig5}(a)).

To better understand the spectral features, higher-resolution spectra were also measured at  $\theta=0^{\circ}$ and $\theta=90^{\circ}$, that is, with light polarization parallel and perpendicular to the GNR edges, respectively (Figs.~\ref{fig:fig5}(c,d)), revealing the light polarization dependence of the Raman spectra of the facet-graphene. The experimental spectra were fitted with six Lorentzian peaks, which were categorized into three origins: GNRs, BLRs, and an unknown component. The peaks having BLR origin were identified since their positions approximately matched the peaks in the 6R3 buffer layer on (0001) macroterraces observed in the blue spectrum in Fig.~\ref{fig:fig5}(a); they were attributed to a carbon sheet with $sp^3$ bondings of BLRs (similarly to the 6R3 structure). The origin of the yellow peak at ${\sim}1500$~cm$^{-1}$ was difficult to identify, but it may be associated with the edge-mode \cite{Ra-b3}. The curve with GNR origin consisted of three peaks: one $G$ peak at ${\sim}1610$~cm$^{-1}$ and two $D$ peaks at ${\sim}1370$~cm$^{-1}$. The latter ones were probably due to either the bulk- and edge-phonon contributions or the phonon energy difference depending on the scattering schemes for the $D$-band transition. The $G$ peak was at a higher wavenumber compared with that for pristine graphene (${\sim}1580$~cm$^{-1}$) reported by Lee \textit{et al}.\cite{Ra-d1} We could not determine the cause of this blue shift, but it may be due to compressive strain, hole doping, or band gap opening\cite{Ra-d1,Ra-d2,Ra-d3}. Band gap opening is the most plausible cause since our GNRs were $2$~nm wide and of the armchair-edge type, which results in a band gap larger than $0.4$~eV, based on the calculations  \cite{Ra-d4}.

Finally, the polarization dependence shown in Fig.~\ref{fig:fig5}(c,d) indicates that the spectra at each angle were similar in shape but an intensity reduced by $1/4.5$. This is a typical feature of armchair-edge graphene\cite{Ra-b1,Ra-b2}. All the Raman spectroscopy results described above confirm the formation of quasi GNR arrays on the macrofacets.

\section{Conclusions}\label{sec:concl}
We studied graphitization phenomena on the macrofacets formed during thermal decomposition of vicinal $6H$-SiC(0001) surfaces with a relatively large miscut angle (${\sim}15^{\circ}$) toward the $[1\bar{1}00]$ direction. The macrofacets were inclined by $27^{\circ}$ with respect to the (0001) macroterraces because of macrostep bunching and graphitization. Monolayer carbon was formed on the macrofacet surface and sectioned by GNR and BNR areas due to periodically striped surface structures. An array of aligned GNRs, each one having a width of ${\sim}2$~nm and being lined up with a $3.3$~nm period, was identified.

The Raman spectrum of facet-graphene revealed the polarization dependence of the $G$ and $D$ peaks and the absence of the $2D$ peak, confirming the existence of armchair-edge GNRs with narrow width. The GNRs were separated by BLRs, which implies that they were electronically isolated. The growth of a uniform pseudo GNR array on a semiconductor substrate via simple annealing is promising for applications in next-generation electronics.

\end{document}